# Enhanced ULF electromagnetic activity detected by DEMETER above seismogenic regions

M. A. Athanasiou[1,2], G. G. Machairidis[3], C. N. David[2] and G. C. Anagnostopoulos[4]

1. Dept. of Information & Communications, Technical University of Serres, Greece
2. Dept. of Mechanical Engineering, Technical University of Serres, Greece
3. Connect I.T, Informatics Company, Greece
4. Dept. of Electrical & Computer Engineering, Democritus University of Thrace, Greece

Correspondence and requests for materials should be addressed to M.A. Athanasiou (Email: mathanas@ee.duth.gr).

**Abstract**
In this paper we present results of a comparison between ultra low frequency (ULF) electromagnetic (EM) radiation, recorded by an electric field instrument (ICE) onboard the satellite DEMETER in the topside ionosphere, and the seismicity of regions with high and lower seismic activity. In particular, we evaluated the energy variations of the ULF Ez-electric field component during a period of four years (2006-2009), in order to examine the possible relation of ULF EM radiation with seismogenic regions located in Central America, Indonesia, Eastern Mediterranean Basin and Greece. As a tool for evaluating the ULF Ez energy variations we used Singular Spectrum Analysis (SSA) techniques. The results of our analysis clearly show a significant increase of the ULF EM energy emitted from regions of highest seismic activity at the tectonic plates boundaries. We interpret these results as suggesting that the highest ULF EM energy detected in the topside ionosphere is originated from seismic processes within Earth's crust. We understand the results of the present study as confirming previous evidence that the ULF EM anomalous wave activity in the ionosphere can be considered a useful earthquake precursory signal.

## 1 Introduction
Although Earthquakes (EQs) are known to be complex phenomena, there is growing evidence in the last decades that EQs precursory phenomena can be detected. This evidence was based on studies of certain effects related to electric and magnetic fields, ionospheric perturbations, nightglow observations, electromagnetic (EM) emissions from DC to high frequency (HF) range and radiation belt electron precipitation in the topside ionosphere (Varotsos et al., 1993; Pulinets, 2011; Hayakawa et al. 2010; Parrot, 1994 ; Anagnostopoulos et al., 2012)

Theoretical studies and laboratory experiments suggest two main mechanisms for the production of precursor earthquake waves, namely the electromechanical mechanism and the acoustic mechanism. These mechanisms are mainly based on the deformations of rocks under special pressure and temperature conditions existing in the brittle seismogenic crust, which destabilise the mechanical and electrical properties of the solids. In particular, according to the electromechanical mechanism, electric charges are generated as the result of friction and piezoelectric phenomenon that changes the Earth's electric field and generates EM waves, which are considered to propagate to the upper atmosphere and ionosphere (Cress et al., 1987; Enomoto et al., 1990; Parrot et al., 1993).

On the other hand according to the acoustic mechanism, gravity waves are generated before and after the earthquake. These waves propagate in the atmosphere, where their amplitudes are increased relatively to height, due to the air's density decrease, disturb the ionosphere and cause the VLF emissions of EM waves (Hayakawa, 2004).

From the 80's until now, dozens of studies regarding several seismic events have been presented validating the existence of EM wave emission during seismic activity (Parrot et al., 2006; Athanasiou et al., 2011) . Nevertheless, bearing in mind the relatively few statistical analyses have been done, the corresponding results still seem to be ambiguous. Only a part of them have shown a temporal and spatial correlation between the EM waves and the seismic events (Parrot, 1994; Henderson et al.,1993; Nemec et al., 2009). So, the question whether there is a relationship between the seismic activity of an area and the EM radiation in the space over this specific area is still open (Masci, 2006). In this research work we are going to further elaborate the above question.

Specifically, instead of studying the ULF EM wave activity around the time of an earthquake occurrence near its epicentre, we compare the ULF EM wave energy (higher than $0.02$ $(mV/m)^2$ ) in the topside ionosphere, with the seismic activity of various geographic regions of historically high seismicity including boundaries of tectonic plates. To this purpose, we analyze ULF data recorded by the ICE sensors onboard the satellite DEMETER (Detection of Electromagnetic Emissions Transmitted from Earthquake Regions) , (Berthelier, et al., 2006). This microsatellite was launched on June 29th, 2004. Its orbit altitude is 710 km, and it takes 14 orbits per day around the Earth. Among the scientific objectives of the DEMETER mission is the investigation of the Earth Ionosphere, disturbances due to seismic and volcanic activities. Our study is focusing on certain regions of great seismologic interest, as Eastern Mediterranean Sea, Central America, Indonesia and Greece. The results strongly support the correlation between ULF EM radiation and the seismic activity of highly seismogenic regions, and particularly the tectonic plates boundaries.

## 2 Methodology and Results

We estimated the energy of the electric field Z-component of ULF electromagnetic waves radiated within the frequency range 0 to 20 Hz as they recorded by the ICE experiment for a period of four years (2006-2009) during the satellite overnight passes above the regions we studied in this research work. The energy was estimated in successive orbit sections with a latitude range equal to $2^0$, under the condition that the middle of these sections is located within the boundaries of the regions we wanted to study. In order to estimate the EM energy, we developed a novel method that comprises two stages: first we filter the signal and we keep only the very low frequencies and afterwards we eliminate its trend using techniques of Singular Spectrum Analysis (SSA), combined with a third-degree polynomial filter (Athanasiou et al., 2001; 2011) Electromagnetic energy less than $0.02$ $(mV/m)^2$ it was not considered in our investigation because it is not significant and has minor impact on the results.

## 2.1 The case of seismicity in Greece

The outstanding seismicity of Greece (the country is ranked third worldwide and first among the countries of the European continent) is due to its special geological features, which are shaped by the movements of Eurasian, African and Anatolia tectonic plates in the Eastern Mediterranean basin. This region of high seismicity is known as the "Greek Arc" and is shown in Figure 1a (the seismic map has been constructed by the University of Athens for earthquakes with magnitude M≥4 of Richter scale occurred in the years 2000-2008). Here we examine the north-west part of the "Greek Arc", a rectangular region with latitudes between $N36^0$ - $N39^0$ and longitudes $E19^0$ - $E22^0$ (seismic region marked as R1). As we can see in Figure 1a, the region R1 (marked by red border line) shows stronger seismic activity, between year 2000 and 2008, than its eastern neighbouring region R2 (marked by green border line). Given that both of the areas (R1 and R2) have same latitude, they absorb comparable solar energy and have equivalent geomagnetic conditions.

During a period of four years (2006-2009) a sample of 41 ULF events with energy $E \geq 0.02$ $(mV/m)^2$ were recorded above both regions R1 and R2, which are shown by solid circles in Figure 1b. The size of the cycles shown in the figure compared with the numeric scale of the label inserted at the bottom of the figure indicates the amount of energy of the corresponding ULF electromagnetic radiation. The number of the events included in Figure 1b corresponds to 9.55% of the total number (430 records) of the ULF events selected according to the described methodology.

From the sample of 41 ULF events, a number of 28 events were recorded above region R1 of historically high seismicity, whereas only 13 events were detected above the adjacent region R2 of low seismicity. The corresponding total emitted energy from the two examined regions was $E_1$= 1.77 $(mV/m)^2$ and $E_2$=0.58 $(mV/m)^2$ respectively. Hence the ratio of the emitted energy from R1 and R2 is $E_1 / E_2$ =3.05. Table 1 shows the values of the total emitted ULF energy from regions R1 and R2 and for the years 2006-2009, as well as per year. As we can see the emitted energy was permanently higher above R1 than above R2 for each year, except for year 2009 when the ULF energy was very low for both areas.

In a next step, assuming that the seismic activity is source of the electromagnetic emission and in order to describe more accurately the relationship between the recorded EM emission and the seismicity of the examined regions, we studied a smaller sample characterized by the most intense ULF events. Thereby, from the initial sample we rejected the records with energy lower than 0.08 $(mV/m)^2$. So, only 9 ULF events remained (~2.09% of the total number of events). From these, 7 events are related with the region R1, while only 2 events with the region R2. In addition we found that the total emitted energy from the two regions amounts $E_1$= 0.96 $(mV/m)^2$ and $E_2$= 0.22 $(mV/m)^2$ respectively. These values give a ratio $E_1 / E_2$ = 4.36, which compared to the ratio $E_1 / E_2$ = 3.05 (sample of 41 events), emphasize the ULF EM energy amount emitted from the two areas (R1 and R2) in relation to their seismicity.

Figures 1c and 1d illustrate the total energy of both considered areas in relation to longitude by means of normalized histograms. It is obvious that the total energy over the area R1, which is characterized by higher seismic activity, is greater than the corresponding energy emitted from area R2 with lower seismicity.

**2.2 The case of seismicity in Central America**
The next area we have studied covers part of Central America and the Pacific Ocean within the latitude range $N10^o$ to $N20^o$ and the longitude range $E215^0$ to $E275^0$. In Figure 2a is shown the seismicity map of the above region for earthquakes with magnitude M ≥ 4.5 of Richter scale and for the period 2000-2009 (sources: University of Athens and USGS). As it is depicted in the map, the eastern region enclosed with the red border line (marked as R3) has more intense seismic activity in comparison to the western region indicated by green border line (marked as R4).

For a period of four years (2006-2009) a sample of 605 ULF events with energy $E \geq 0.02$ $(mV/m)^2$ were recorded by the satellite DEMETER during its night passes over the examined areas (this sample corresponds to 4.22% of the ULF events totally recorded). From this sample, 416 events were recorded above region R3, whereas only 189 events were detected above the region R4, while the corresponding total emitted energy was evaluated to be $E_3 = 19.91$ $(mV/m)^2$ and $E_4 = 7.96$ $(mV/m)^2$, espectively. Representative results of the geographic distribution of the ULF events within the year 2008 are shown in Figure 2b. In the same figure the seismic fault lines of the region due to the movements of the tectonic plates (Pacific Plate, Cocos Plate, Caribbean Plate and North American Plate) are illustrated. It is obvious that a similar asymmetry in the distribution of the ULF events between areas R3 and R4 for the year 2008 was observed, as in the case during the whole period 2006-2009 discussed above. More explicitly, we found 108 events above the region R3 (total energy $E_3 = 5.44$ $(mV/m)^2$), but only 46 events above the region R4 ($E_4 = 1.76$ $(mV/m)^2$). In addition, it is worth noting that the density of the electromagnetic energy increases close to the fault branches of this seismogenic region.

Table 2 shows the ULF EM energy values regarding the examined regions R3 and R4 for the years 2006-2009 as well as per year. As we can see the total energy of the recorded ULF events during the whole period, as well as for each year is greater above region R3 (with higher seismic activity) than above region R4. Specifically, the ratio of the emitted energy from the regions R3 and R4 for the years 2006-2009 was found to be $E_3 / E_4 = 2.5$.

Based on the methodology explained in the case of Greece, when only the most intense ($E \geq 0.08$ $(mV/m)^2$) ULF events are taken into account (52 events, i.e. 0.36% of the total sample), an asymmetrical distribution regarding both the event numbers ($N_3$, $N_4$) and the total emitted energy ($E_3$, $E_4$) for the regions R3 and R4 ($E3 / E4 = 3.7$) was found. Thereafter $N_i$ is considered as the number of ULF events and $E_i$ as the emitted ULF EM energy in the corresponding region $R_i$. (In this case $N_3 = 42$, $N_4 = 10$, $E_3 = 5.47$ $(mV/m)^2$ and $E_4 = 1.48$ $(mV/m)^2$). This asymmetry is similar as in the case we examined above ($E \geq 0.02$ $(mV/m)^2$), but with a higher ratio of the emitted energies.

Figures 2c and 2d show the normalized histograms of the total energy of both considered regions. It is obvious that the total energy over the area R3, which is characterized by higher seismic activity, is greater than the corresponding energy emitted from area R4 with lower seismic activity.

**2.3 The case of seismicity in Indonesia**
Figure 3 has been constructed in the same format as Figures 1 and 2 concerning a large region of Indonesia, between Lat. $0^0$ - $S10^0$ and Long. $E80^0$ - $E110^0$ and for earthquakes of magnitude $M \geq 4.5$ of Richter scale for the years 2000-2009. As we can see (Fig. 3a) the seismic activity is much higher in the region R5 ($E95^0$ - $E110^0$) compared to the region R6 ($E80^0$ to $E95^0$) due to the complex of the boundaries of the Australian-Indian and Eurasian tectonic plates. In this case 245 ULF events with energy $E \geq 0.02$ $(mV/m)^2$ were selected by our algorithm (i.e. 3.4% of the total ULF events). Among them 156 recorded above region R5 (total emitted energy $E_5 = 7.22$ $(mV/m)^2$) and only 89 events detected above region R6 ($E_6 = 3.5$ $(mV/m)^2$) respectively.

Figure 3b demonstrates a representative sample of the geographical distribution of the recorded ULF events for the year 2006. We found 44 events above the region R5 (total energy $E_5 = 2.19$ $(mV/m)^2$), but only 22 events above region R6 ($E_6 = 0.65$ $(mV/m)^2$). It is obvious that the density of the ULF events and the total emitted energy apparently becomes higher over the main fault, which is depicted in blue colour line.

Table 3 shows similar results with Tables 1 and 2 concerning the regions R5 and R6. For the whole period 2006-2009 the ratio of the emitted energy from R5 and R6 was found to be $E_3 / E_4 = 2.06$. As in the previous cases (Fig. 1d and Fig. 2d), when only a set with the most intense ($E \geq 0.08$ $(mV/m)^2$) ULF events are considered (0.30% of the total sample), a similar asymmetry of the emitted energy distribution is found, as for the events with energy $E \geq 0.02$ $(mV/m)^2$, but with a higher ratio ($E_5 / E_6 = 5.18$). The numbers of the considered ULF events and the total emitted energies were $N_5 = 18$, $N_6 = 4$, $E_5 = 1.97$ $(mV/m)^2$ and $E_6 = 0.38$ $(mV/m)^2$ respectively. Figures 3c and 3d reveal a similar pattern for the electromagnetic energy emitted from the more seismically active region compared to the cases that have been presented in Figures 1 and 2.

Since the above examined regions (R5 and R6) lie at one of the broader and most seismically active areas of the globe, in order to further prove our results we extended our study on a larger longitude range including the whole seismic region between Indonesia and Australia. For this reason we analyzed data for the region between Lat. $0^0$ - $S10^0$ and Long. $E80^0$ - $E180^0$. This enlarged region (named R7) includes a wide part of the boundaries of the Australian plate with the Eurasian and the Pacific plates of highest seismicity (Fig. 4a; red coloured frame). The left and the right sub-regions of the drawn frame are characterized by lower seismic activity. Furthermore, for a detailed analysis, we divided the whole region into 10 geographical zones, with a longitude range of $10^0$ each one. For the years 2006-2009, 171 events with energy $E \geq 0.06$ $(mV/m)^2$ were found and the total emitted energy ($E_7$) amounts 14.36 $(mV/m)^2$.

Table 4 shows the number of ULF events over each geographical zone and the coresponding total emitted ULF electromagnetic energy. For facilitating a direct comparison of the emitted ULF wave energy with the seismic activity of the region R7, Figure 4b shows a histogram of the total energy, normalized at its maximum value for the 10 sub-regions. By comparing Figures 4a and 4b we can see a very good correlation between the density of large EQs and the total emitted ULF electromagnetic energy $E_{7-i}$, where the index i corresponds to the examined sub-regions ($E80^0$ - $E90^0$, $E90^0$ - $E100^0$ ... $E170^0$ - $E180^0$). Additionally, in the zone within the longitudes $E120^0$ - $E130^0$ the total energy becomes lower, which is possibly related with the fault disruption as it is indicating by the blue line. The results are more evident if waves with total energy less than 0.12 $(mV/m)^2$ are rejected, as shown in Table 4 and graphically illustrated in Figure 4c. In this case the total wave energy takes zero values over the zones with longitudes $E80^0$ - $E90^0$ and $E160^0$ - $E180^0$, while the energy decreasing in the zone with longitudes $E120^0$ - $E130^0$ becomes more obvious.

**2.4 The case of seismicity in Eastern Mediterranean Basin**
In our analysis we have presented a comparison of ULF electromagnetic wave activity above adjacent regions in three examples of highly seismic areas on the globe: Greece, Central America and Indonesia. More explicitly, we compared the ULF electromagnetic wave activity in two distinct regions in each of the three cases, with the one known as highly seismogenic and the adjacent one with low or very low seismic activity. Moreover, we have also mentioned that the ULF electromagnetic wave activity was higher in the vicinity of the earthquake faults of the regions examined. In this section we would like to further elaborate the question how the ULF electromagnetic wave activity varies when the satellite DEMETER passes close to the earthquake faults. As an example we examine the Eastern Mediterranean basin.

Figure 5a shows the geographical distribution of the ULF events with energy E $\geq$ 0.02 (mV/m)$^2$ recorded by the satellite DEMETER during the period 2006-2009 over the Eastern Mediterranean Sea with latitude ranging between N33$^0$– N41$^0$ and longitude E16$^0$ - E36$^0$. In Figure 5a, some concentrations of ULF events are rather evident: southern Ionian Sea, the south part of Crete and the northern border between Greece and Turkey (shaded by red). In this case we want to identify the DEMETER positions in which the highest ULF wave radiation is recorded. For this reason we show in Figure 5b only those events with energy E $\geq$ 0.14 (mV/m)$^2$. The majority of the intense ULF events (yellow colored cycles) is observed very close to the faults (9 over 12 i.e. 75%). This finding demonstrates a strong correlation between the DEMETER positions where the highest ULF wave radiation has been recorded, with the characteristic regions of the well known seismic faults in the Eastern Mediterranean Sea basin.

## 3 Discussion and Conclusions

This paper aims to testify the hypothesis that the ULF wave activity is stronger above regions of high seismicity, and in particular at the boundaries of the tectonic plates, compared to the ULF wave activity detected above adjacent regions. In order to verify this hypothesis the energy of the electric field component Ez of the ULF radiation being recorded by the DEMETER satellite during its night time passages over seismically active regions such in Greece, Central America, Indonesia and Eastern Mediterranean basin for a period of four years (2006-2009) was estimated.

The data analysis of the present study suggests that: (1) the higher the region's seismicity is, the more enhanced electromagnetic radiation is detected, (2) the most intensive electromagnetic ULF wave radiation is recorded during satellite trajectories passing near or just above seismic faults, and (3) the enhanced ULF wave activity above regions of higher seismicity (at the boundary of tectonic plates) compared to the adjacent regions was found to be a permanent phenomenon during four successive years (2006 - 2009).

Nevertheless, someone might attribute the enhanced ULF wave activity to other sources, such as human activities, or to some other natural phenomena, which might have an impact on the ULF EM waves in the topside ionosphere. However, we found that when examining the Greek territory (Fig. 1b), much higher electromagnetic radiation was detected in the region above the Greek Arc (R1) than above the adjacent region (R2), which includes the city of Athens, with about four million residents. Obviously, we see in this case that the human activity related ULF electromagnetic radiation is of minor importance. Furthermore, since we compared regions adjacent the one to each other and located at the same latitudes, we can assume that global or large scale natural phenomena, as for instance, geomagnetic disturbances or solar wind impact most probably cannot explain the systematic local differentiation of ULF radiation near the boundaries of the tectonic plates. This conclusion is greatly supported by the permanent character of the phenomenon we found for a time period as long as 4 years.

Bearing in mind the results that have been presented, we believe that this research work greatly supports the hypothesis that ULF electromagnetic radiation is actually emitted from seismic areas and hence, the ULF EM radiation could be seriously considered as a precursor of earthquakes. Our results are consistent with the theoretical model recently proposed by Molchanov (Molchanov, 2011), as well as with other related studies.

It remains to study in which manner the main characteristics (intensity, frequency etc.) of this kind of electromagnetic waves are modulated during the pre-earthquake period. Such a study might be a very helpful tool to investigate and possibly to predict the time of the main shock of earthquakes. At this point, we would like to mention, that we found a significant increase of the ULF electromagnetic radiation 25 days before the catastrophic Haiti earthquake and a gradual decrease after the main shock (Athanasiou et al., 2011). Similar results came out by studying three other recent earthquakes (Greece, Japan and Italy) and we hope that they will be published soon. Moreover, the implemented methodology is capable to detect even small seismic areas during their pre-earthquake activity, as an example the case of the Greek Arc examined in this study (Fig. 1).

To recapitulate, the present work provides strong evidence that the ULF electromagnetic radiation in the ionosphere can be accepted as a precursor of earthquakes. Our next target is to further develop our methodology, so that we will be able to contribute in the effort of the international scientific community to answer the questions where, when and of which magnitude an earthquake will occur.

**Acknowledgments**
The financial support by the Research Committee of the Technical University (TEI) of Serres under grant 117/13/3.10.2012 is gratefully acknowledged.

Correspondence and requests for materials should be addressed to M.A.
(Email: mathanas@ee.duth.gr; athanasiou@teiser.gr).

| Table 1 Total Energy of the ULF electromagnetic waves expressed in (mV/m)$^2$ | | | | | |
|---|---|---|---|---|---|
| Greece | Years | | | | |
|  | 2006 | 2007 | 2008 | 2009 | 2006-2009 |
| Region R1 Total Energy (ER1) | 0.20 | 0.77 | 0.75 | 0.05 | 1.77 |
| Region R2 Total Energy (ER2) | 0.12 | 0.04 | 0.33 | 0.09 | 0.58 |
| ER1/ER2 | 1.67 | 19.25 | 2.27 | 0.55 | 3.05 |

| Table 2 Total Energy of the ULF electromagnetic waves expressed in (mV/m)$^2$ | | | | | |
|---|---|---|---|---|---|
| Central America | Years | | | | |
| | 2006 | 2007 | 2008 | 2009 | 2006-2009 |
| Region R3<br>Total Energy (ER3) | 3.36 | 3.89 | 5.44 | 7.22 | 19.91 |
| Region R4<br>Total Energy (ER4) | 2.60 | 1.91 | 1.76 | 1.69 | 7.96 |
| ER3/ER4 | 1.29 | 2.03 | 3.09 | 4.27 | 2.5 |

| Table 3 Total Energy of the ULF electromagnetic waves expressed in $(mV/m)^2$ | | | | | |
|---|---|---|---|---|---|
| Indonesia | Years | | | | |
| | 2006 | 2007 | 2008 | 2009 | 2006-2009 |
| Region R5 Total Energy (ER5) | 2.19 | 1.44 | 1.82 | 1.77 | 7.22 |
| Region R6 Total Energy (ER6) | 0.65 | 0.98 | 0.49 | 1.38 | 3.50 |
| ER5/ER6 | 3.36 | 1.47 | 3.71 | 1.28 | 2.06 |

| Longitude | Energy threshold 0.06 $(mV/m)^2$ | | Energy threshold 0.12 $(mV/m)^2$ | |
|---|---|---|---|---|
| | Number of Events | Total Energy | Number of Events | Total Energy |
| E $80^0$-$90^0$ | 10 | 0.77 | 0 | 0 |
| E $90^0$-$100^0$ | 10 | 0.89 | 2 | 0.26 |
| E $100^0$-$110^0$ | 25 | 2.23 | 4 | 0.55 |
| E $110^0$-$120^0$ | 24 | 1.92 | 2 | 0.25 |
| E $120^0$-$130^0$ | 18 | 1.44 | 1 | 0.12 |
| E $130^0$-$140^0$ | 24 | 2.23 | 5 | 0.68 |
| E $140^0$-$150^0$ | 24 | 2.04 | 4 | 0.52 |
| E $150^0$-$160^0$ | 23 | 1.88 | 2 | 0.26 |
| E $160^0$-$170^0$ | 7 | 0.51 | 0 | 0 |
| E $170^0$-$180^0$ | 6 | 0.45 | 0 | 0 |

Table 4 Total energy of the ULF electromagnetic radiation over Indonesia

Latitude: $0^0$-S$10^0$
Years: 2006-2009

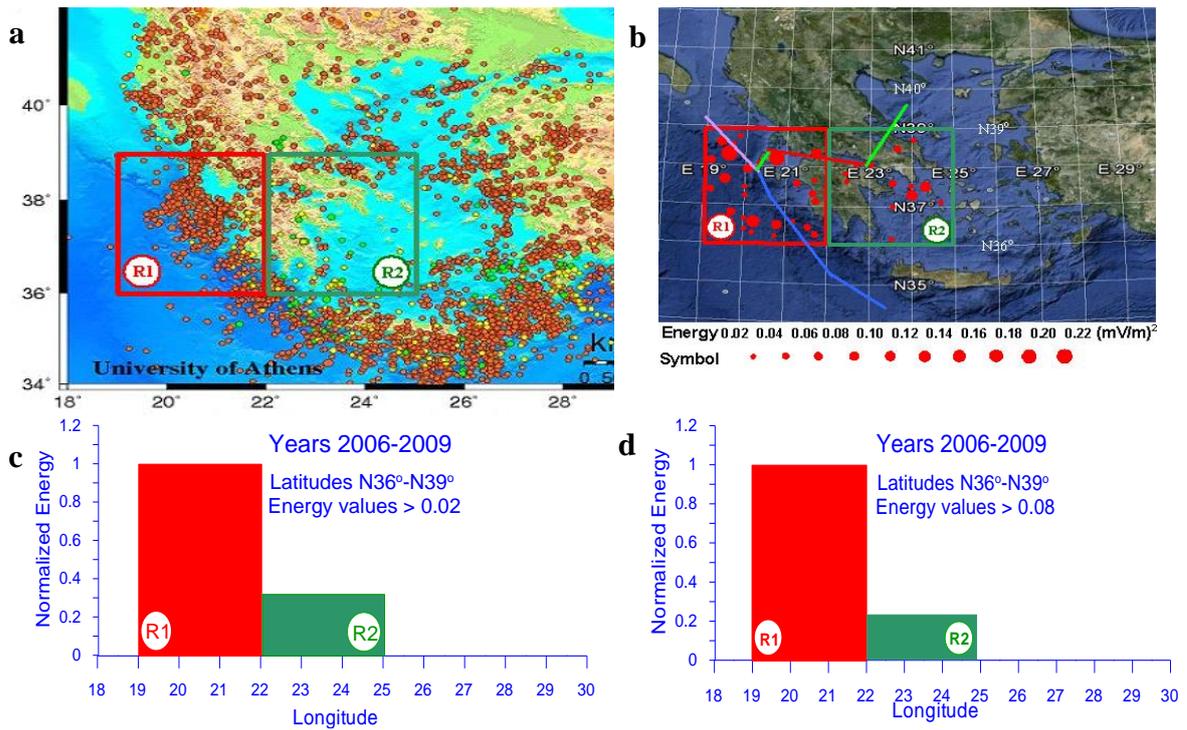

**Fig. 1. (a)** Seismicity map of Greece for the years 2000-2008 illustrating earthquakes of magnitude M>4. **(b)** Location and energy of the ULF electromagnetic radiation acquired by the DEMETER satellite for the years 2006 - 2009. **(c)** Normalized energy of ULF EM radiation in the areas R1, R2 for values greater than 0.02 $(mV/m)^2$. **(d)** Normalized energy of ULF EM radiation in the areas R1, R2 for values greater than 0.08 $(mV/m)^2$.

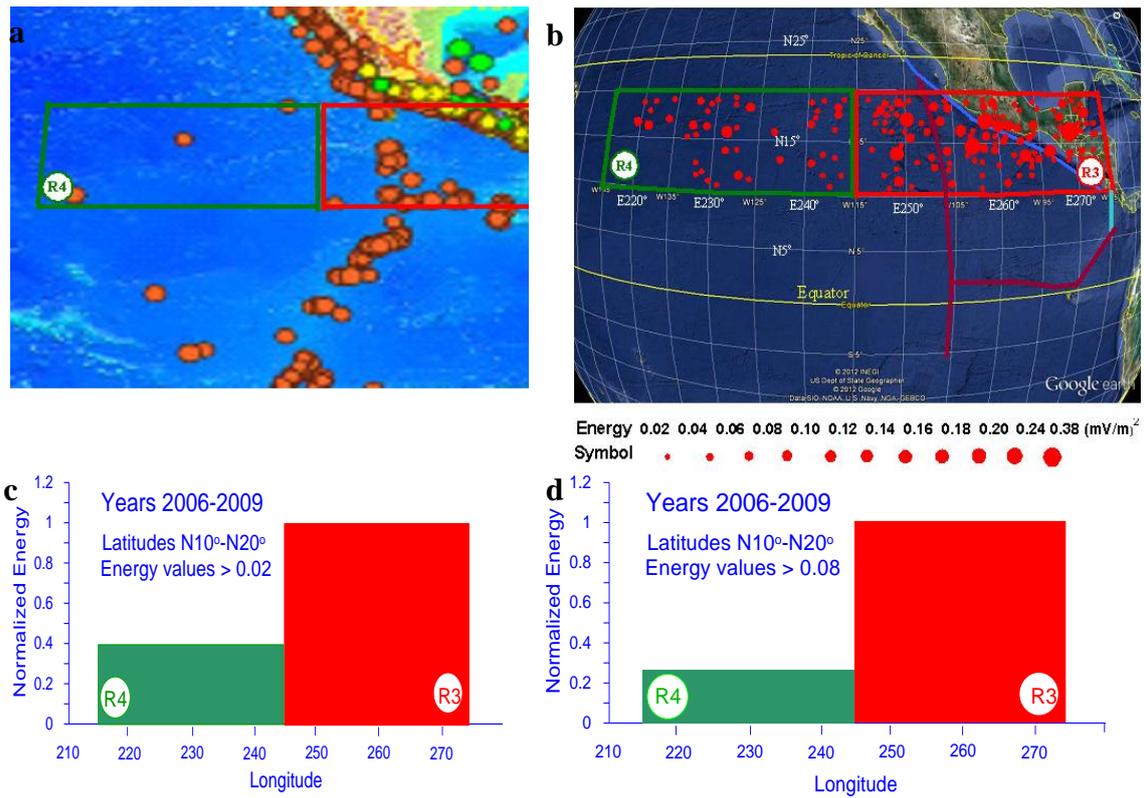

**Fig. 2. (a)** Seismicity map of Central America for the years 2000-2009 illustrating earthquakes of magnitude M>4.5. (**b**) Location and energy of the ULF electromagnetic radiation acquired by the DEMETER satellite for the year 2008. **(c)** Normalized energy of ULF EM radiation in the areas R3, R4 for values greater than 0.02 $(mV/m)^2$. **(d)** Normalized energy of ULF EM radiation in the areas R3, R4 for values greater than 0.08 $(mV/m)^2$.

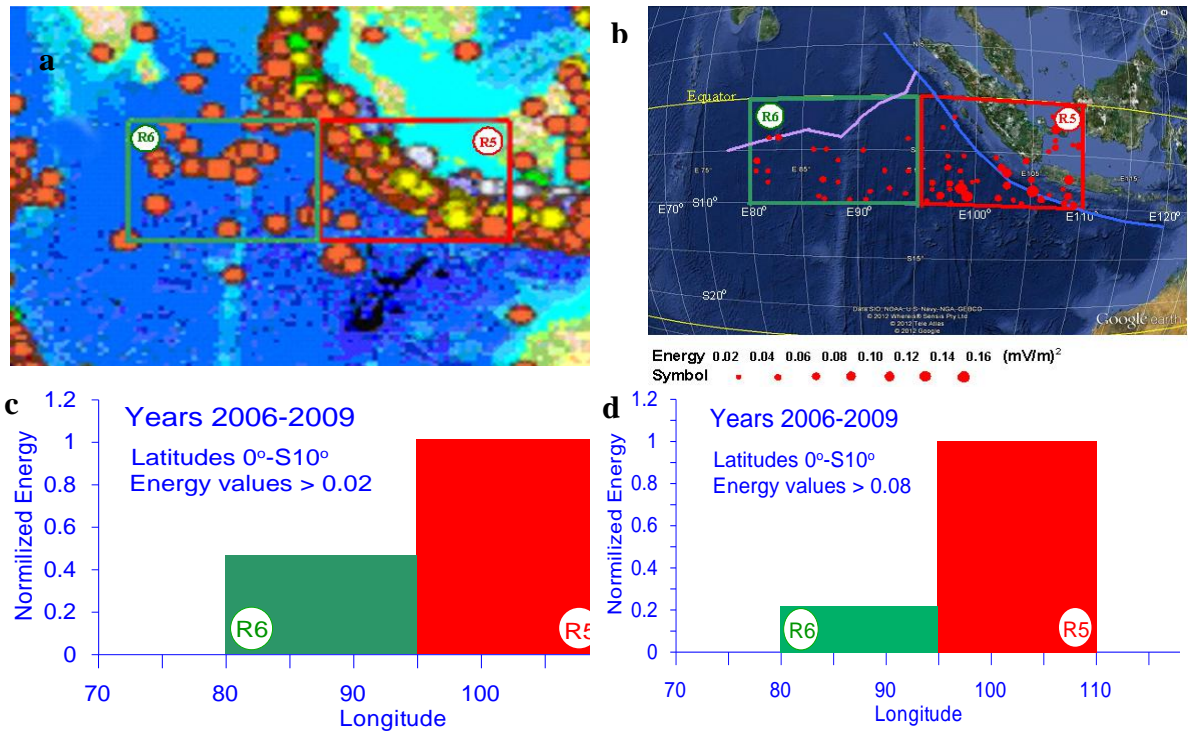

**Fig. 3. (a)** Seismicity map of Indonesia for the years 2000-2009 illustrating earthquakes of magnitude M>4.5. **(b)** Location and energy of the ULF electromagnetic radiation acquired by the DEMETER satellite for the years 2006. **(c)** Normalized energy of ULF EM radiation in the areas R5, R6 for values greater than 0.02 $(mV/m)^2$. **(d)** Normalized energy of ULF EM radiation in the areas R5, R6 for values greater than 0.08 $(mV/m)^2$.

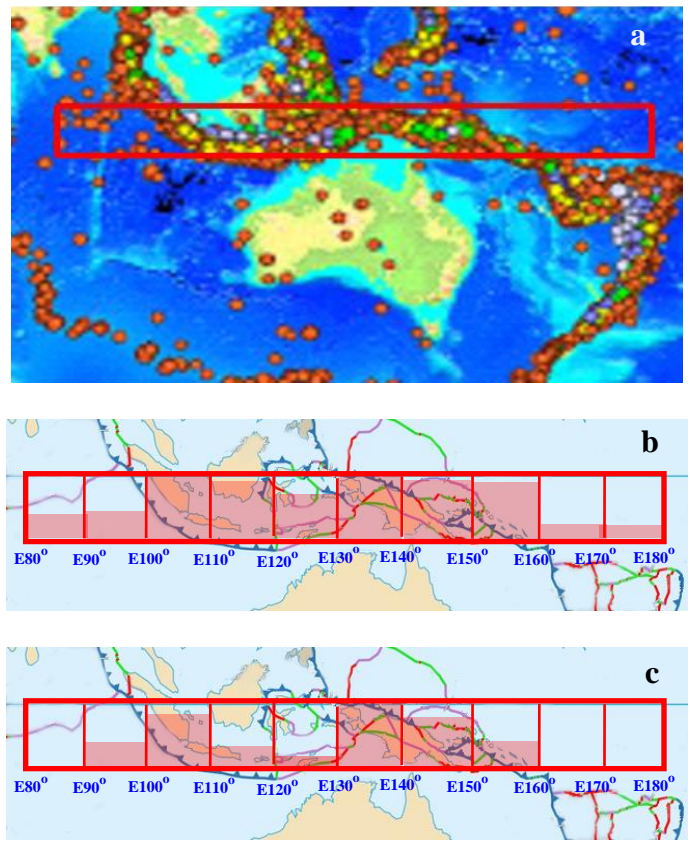

**Fig. 4.** **(a)** Seismicity map of Indonesia for the years 2000-2009 illustrating earthquakes of magnitude M>4.5. **(b)** Normalized energy of ULF EM radiation within $10^0$ longitude divisions acquired by the DEMETER satellite for the years 2006-2009 with values greater than 0.06 $(mV/m)^2$. **(c)** As in subfigure (b) but for values greater than 0.12 $(mV/m)^2$.

**Fig. 5. (a)** Location and energy of the ULF electromagnetic radiation acquired by the DEMETER satellite for the years 2006 - 2009 and for values greater than 0.02 $(mV/m)^2$. **(b)** As in subfigure (a) but for energy values greater than 0.14 $(mV/m)^2$.